4# SUCCESSFUL E-BUSINESS SYSTEMS - PAYPAL

4**Archil Avaliani**

**International University in Germany**

**Supervisor: Prof. Keiichi Nakata**

**ABSTRACT**

PayPal is an account-based system that allows anyone with an email address to send and receive online payments. This service is easy to use for customers. Members can instantaneously send money to anyone. Recipients are informed by email that they have received a payment. PayPal is also available to people in 38 countries. This paper starts with introduction to the company and its services. The information about the history and the current company situation are covered. Later some interesting and different technical issues are discussed. The Paper ends with analysis of the company and several future recommendations.

**TABLE OF CONTENTS**

## 1. INTRODUCTION

### 1.1 - PayPal – A Popular Company

This paper is about a company called PayPal. First I will touch the general information about the company, then provide information about several details and finish with several useful future recommendations.

PayPal (formerly X.com) that was founded in 1998 belongs to an Electronic Payment Systems, more precisely to Generic Systems subgroup. It is located in San Jose and carries out services that provide the possibility of sending and receiving money by means of computers in an easy, secure and fast way using an account-based system. This can be done by anyone (a private person or legal entity) who has an email address. When the money is transferred, a message is sent to the recipient's email address notifying about the transaction. At the same time certain security measures are taken during these operations. One can still charge his or her account without having an electronic card, using the bank transfer or a paycheck. Transferring money is as easy as sending an email for customers.

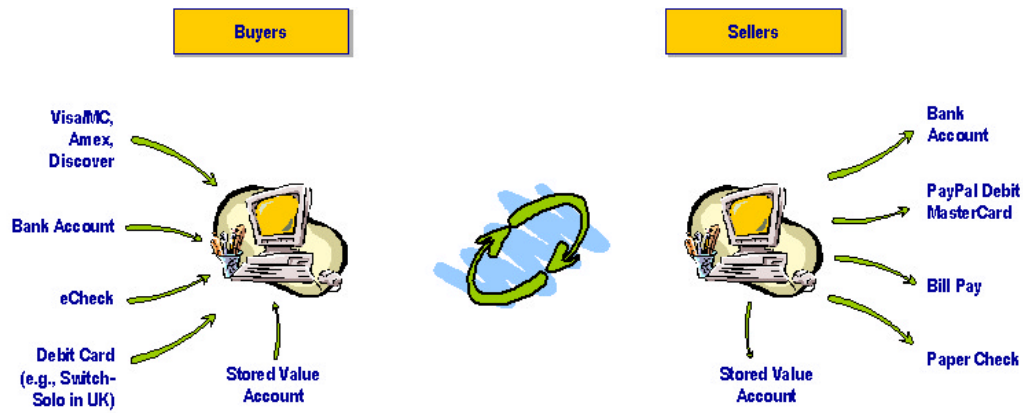

(Fig.1, What is PayPal, retrieved on February 15th, 2004 From: PayPal Developer Conference Keynote, Alex Kazim, VP Marketing, 05/08/03. Data available on http://paypaldev.org/ )

In 2002, PayPal was acquired by eBay. It has quickly become popular. Today the company offers services in thirty-eight countries and has about forty million customers for today. In most of these countries it supports bank withdrawal service and is especially popular among eBay customers. Fig.1 briefly illustrates Company's business. The system enables its users to send money, using computers. The same can be done by means of mobile phones that are support Web. During money transfer the service fee is charged to the customers' bank account, credit card or PayPal balance. At present two thirds of the customers use this service while participating in an online auction. It must also be noted that opening a Personal account and certain basic services provided by PayPal are free of charge. There are certain charges for opening Premier and Business accounts and for additional types of services. The company has a successful business. Number of customers is increasing. Therefore the company's profits also. In year 2002 company had a significant gross. Its one-year sales growth was 134% ($236.6) and one-year employee growth was 94.2% (1200 employees).

## 1.2 - Type of the Business

PayPal is a credit card processor or also called E-Transaction Company that allows businesses and individuals to send and receive money. The Jurisdiction differs according to state and sometimes from country to country. As a financial company it has a statute "Money Transmission Act" (may have different name depending on the state).

## 2. PAYPAL AS E-BUSINESS

### 2.1 - Services and Products

Several types of products and services are provided by the company. Individual/Business accounts, Merchant and Auction tools, Third-party software, Shipping services via UPS and a PayPal debit card.

**PayPal Accounts:**

There are three different account types that differ in the services offered. Personal accounts are free, for individual use only, and include only core features. Premier and Business accounts are charged a decent fee to receive payments, and include our premium features, suchlike the ability to accept credit card payments and the PayPal Shopping Cart.

Personal account basically only offers to send money and receive money from non-credit card payments for free. This type of account is especially popular because of its "Auction Tools". Premier features in addition to personals' one are the following. It allow to receive money from credit cars for low fees, has customer service hotline support, offers special seller tools and ATM debit card for selected users. This account type also allows mass payment, which means one can pay thousands of customers at once. Business accounts is the most powerful one. In addition to all services mentioned above it allows to do business under a corporate or group name. Multiple Login is also supported. The following table (Fig.3) clearly shows the difference between account fees.

|  | Personal Account | Premier/Business Account |
|---|---|---|
| Open an Account | Free | Free |
| Send Money | Free | Free |
| Withdraw Funds | Free for US bank accounts<br>Fees for other banks | Free for US bank accounts<br>Fees for other banks |
| Add Funds | Free | Free |
| Receive Funds | Free | 2.2% + $0.30 USD to 2.9% + $0.30 USD† |
| Multiple Currency Transactions | Exchange rate includes a 2.5% fee* | Exchange rate includes a 2.5% fee* |

(Fig.3, PayPal Fees, Retrieved on February 10$^{th}$, 2004 From: http://www.PayPal.com/cgi-bin/webscr?cmd=_display-fees-outside)

### Merchant and Auction Tools

Merchant Tools are for website payments and consequently gives opportunity to accept credit cards for online payments. It allows the user to manage and track his or her payments. For creating the website one can purchase third party tools, that are PayPal products, like Shopping Cart and Buy Now buttons. PayPal also provides other third party tool for making your work easy and also supports you to establish a proper payment process. Several businesses auction tools provided by PayPal are very useful. They helps to manage the whole process for auction. For both Merchant and Auction Tools there is some source code available for free from other developers. PayPal provides developers website and discussion forums where one can get help about development issues. Business customers can add conventional shopping card functionality that is already provided by PayPal.

### Mass Payments:

Some businesses need to make high volume payments to their customers or suppliers. For this case, Mass Pay functionality is provided. One can just upload a text file of special format that includes information about the amount of money and recipients information and PayPal will batch this payments and process them automatically. It is important to note that PayPals Merchant tools support multi-currency property.

### What is PayPal debit card about?

One can request a PayPal ATM/Debit Card and use it to make payments everywhere the MasterCard or Cirrus logos are, or even withdraw the money from an ATM. Unfortunately, as mentioned already, the Debit Card is only available to a limited number of Verified Canadian members at this time.

### 2.2 - Profitable Business

The company gets most of its money from the services and development tools. Because it is considerably cheaper in comparison to other money transfer services and other credit card companies' merchant service, even small businesses can afford it. There are about 42,000 web sites that accept PayPal. The very important thing that made PayPal more profitable, successful and popular company is that it became an eBay Company. As one can notice, Fig.4 and Fig.5 show significant jump in the forth quarter of year 2002. (Q402). Because eBay was already very popular by that time, many customers found it more convenient to use PayPal. This increases the customer number, because eBay was able to make money transactions easily also for customers who didn't have a credit card.

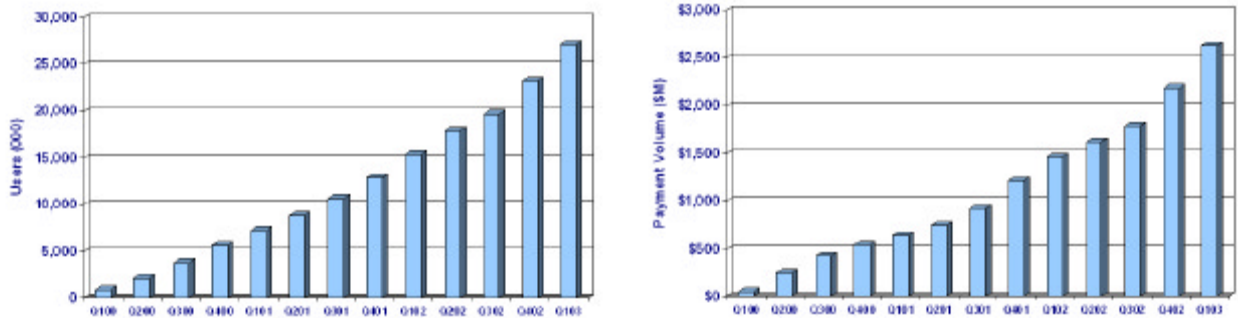

(Fig.4 and Fig.5, Total Payment Volume, retrieved on February 2004 From: PayPal Developer Conference Keynote,Alex Kazim, VP Marketing, 05/08/03. Data available on http://paypaldev.org/)

## 3. TECHNICAL ISSUES

### 3.1 - Security and Scalability of online payments

PayPal consists of four major network components that reside at: PayPal Facilities, Equinix Data Center, Cable and Wireless data center and PayPal Operations and Customer Support facility. The most responsive data and hardware of the network are distributed on Equinix and Cable & Wireless data centers. (eBay Information book)

To reduce the number of failures the company has redundant Internet connections, fault tolerant power and fire control systems and physically secure place. Security is a very important issue for companies like PayPal. Especially because many people try to get money by using the weaknesses of e-business systems. PayPal uses multiple layers of network security; encrypted database and communication with SSL among the systems components to make sure that they provide a reliable service and their customers are secure. Securing IPN (Instant Payment Notification) allows the users to check payment status. The status illustrates the situation of the transaction (Completed, Pending, Canceled, Failed). This allows tracking each transaction and controlling the account. It is especially important for business account users.

However in order to provide significant proofs of customer's identity, it is significant to have a registration form that asks for all necessary information. It is a type of a fraud when third party obtains the PayPal's login information of a consumer and uses it illegally. "If the interloper draws directly on the P2P account, Regulation E makes the P2P intermediary directly responsible". "P2P provider can not charge the consumer's account for the transaction." (Mann, 2003, p.380). There are many different regulations and details that give a lot of information about how to administer transactions. It is strongly recommended for the users to introduce themselves to the regulations of the email money transaction service they use, before they do any serious transactions.

## 4. ANALYSES

### 4.1 - Why did eBay acquire PayPal?

"The success of eBay's auction business had a rare effect of creating a vast market for an entirely new payment product, one that would allow non-merchants (who cannot accept conventional credit-car payments) to receive rapid payments in remote transactions". (Mann, 2003, p.377).

A company like PayPal can provide a lot of advantages to eBay. Not all the people who want to use eBay service can pay using online payments. eBay and PayPal provde an end-to-end e-Commerce solution. Both buyers and sellers on the auction can make money transactions easily.

The main reasons why PayPal was chosen by eBay are the following:

- Security: One of the safest e-commerce platforms.
- Direct Perception
- High Applicability
- Low costs for merchant accounts.
- Flexibility and ease to use: accepts major credit and debit cards as well as direct transfers.
- Scalability (no installation costs necessary)
- Ease of Use

Applicability defines as the degree to which the payment system is acknowledged for different payments. Payments need to be done in an easy way.

Today PayPal continues to be a successful business. It completes over three hundred thousand transactions per day for over forty million accounts. That results into over eighteen million dollars in payments per day. It is important to note that PayPal had incredible growth during the times when most of its competitors failed.

The security polices that PayPal provides are customer oriented and match nicely to the auction business requirements. PayPal's fraud rate is also very important for eBay. In certain cases the system provides for the possibility of protecting the buyer's interests e.g. when a seller is cheating and not sending the purchased goods to the buyer. In such cases everything depends upon the ways of financing the purchase: whether the money is transferred from the buyer's account or (with P2P provider) or credit card. If P2P provider funds the transaction, buyer has no right of asking about the recovering of funds (EFT regime). Only in this last case the buyer has a right to receive the paid amount back if the purchased goods have not been delivered (TILA/Z regime). It is interesting to see PayPals success before and after eBay acquisition. PayPals' revenues increased significantly in the last quarter of the year when e-Bay acquired them. (PayPal Revenues, retrived on March 13th, 2004 from: http://www.shareholder.com/ebay/downloads/pres-feb13.pdf)

## 4.2 - Why companies like PayPal are popular

As mentioned above, PayPal was founded in 1998. It was one of the first successful electronic money transaction companies. (Other credit-debit payment model systems are NetBill, First Virtual's InfoCommerce system, NetCheque, FirePay and more). Later on other companies begin to appear like NOCHEX and iKobo.

The business is popular because it is customer friendly. By providing fast and flexible services to both B2C and C2C money transactions, for cheap prices, the number of people who use this money transaction service growth. "The pervasive regulatory supervision of banks ensures that they honor their (P2P providers) obligations under a variety of consumer-protection and data privacy regulations that govern their activities" (Mann, 2003, p.376).

Many people participate in online gambling. FirePay is a most wildly accepted company among online casinos. Both PayPal and FirePay accounts can be used for online gambling. These accounts become very useful when one needs to use them for participating in an auction or for online gambling, because they allow hiding your credit card information and therefore sending and receiving money easily and securely.

## 4.3 Further Recommendations

The number of companies like PayPal is getting larger. The market gets more competitive. Therefore PayPal needs to find new ways for better customer satisfaction and offering bigger variety of services. Today the company is most popular because it is an eBay Company. This gives a chance to PayPal to invest for company's future.

Company needs to improve the security not only from the technical but also from the company policy's point of view. It provides the service for good prices but it is necessary to improve customer security in order to attract more customers and businesses. Usually it is very difficult to find the optimal regulation that will fit both for the company and its customers.

The issue of whether these kinds of companies need regulation and whether a regulatory body should exist is a complicated one. On one hand it would be natural to regulate the activities of these service providers to guarantee better protection of customers. It is impossible to leave this sphere totally without any regulation leverage. On the other hand the activity of these companies is connected with banks and to a certain extent controlled by them. Additional regulatory leverage might limit banks' activities and impose additional control on them too, which is undesirable.

One last future step is to improve the system design from users point of view. The testing of the new system model with users' involvement increases the chance of finding valid design statements. Therefore the system will get better customer acceptance.

## 5. CONCLUSION

Email money transaction systems become available because it gave possibility to offer money transactions for very cheap prices, to allow the small offline companies to make there business online and to make C2C transactions to become achievable. As it was mentioned in the paper, PayPal started as a small company, and because it had attractive security policies and successful business, the company was acquired by eBay. Today this is good chance and very important opportunity for making the business even more successful in the future.

Low prices and comfortable service is very important for to attract and satisfy customers. PayPal makes main profits from money transactions and merchant services. Even though the transaction fees are low the income is still high, because the company successfully completes over three hundred thousand transactions per day.

The security level of the service offered is high, but like PayPal most of financial companies need to improve their security continuously in order to make accounts safer. The important advantages of the credit-debit model systems are ease of use and scalability. Even though company's customer number increases, there is very unimportant decrease of performance. Direct Perception is PayPal's important advantage.